\documentclass[preprint,aps,amsmath,superscriptaddress,nofootinbib]{revtex4}
\usepackage{bm}
\usepackage{epsfig}
\usepackage{epstopdf}
\usepackage{graphics,graphicx}
\usepackage{fancyhdr,fancybox}
\usepackage{feynmf}
\usepackage[small]{caption}
\usepackage{enumerate}

\def\bsigma{\mbox{\boldmath $\sigma$}}

\newcommand{\nn}{\nonumber} 

\newcommand{\bea}{\begin{eqnarray}}
\newcommand{\eea}{\end{eqnarray}}

\DeclareMathOperator{\tr}{tr}
\newcommand{\lrbd}{\overset{\leftrightarrow}{\partial}}

\newcommand{\bfA}{{\bf A}}
\newcommand{\bfB}{{\bf B}}

\newcommand{\calL}{\mathcal{L}}
\newcommand{\calM}{\mathcal{M}}
\renewcommand{\d}{\partial}

\begin{document}



\title{Line shapes in $\Upsilon(5S) \to B^{(*)} \bar{B}^{(*)}\pi$ with  $Z(10610)$ and $Z(10650)$ using effective field theory}

\author{Thomas Mehen\footnote{Electronic address: mehen@phy.duke.edu}}
\affiliation{Department of Physics, 
	Duke University, Durham,  
	NC 27708\vspace{0.2cm}}

\author{Joshua W. Powell\footnote{Electronic address:    jwp14@phy.duke.edu}}
\affiliation{Department of Physics, 
	Duke University, Durham,  
	NC 27708\vspace{0.2cm}}

\date{\today\\ \vspace{1cm} }


\begin{abstract}

The Belle collaboration recently discovered two resonances, $Z_b(10610)$ and $Z_b(10650)$ --- denoted $Z_b$ and $Z_b^\prime$ --- in the decays $\Upsilon(5S) \to \Upsilon(nS) \pi^+ \pi^-$ for $n$ = 1, 2, or 3, and $\Upsilon(5S) \to h_b(mP) \pi^+ \pi^-$ for $m = 1$ or 2.  These resonances lie very close to the $B^* \bar{B}$ and $B^* \bar{B}^*$ thresholds, respectively.  A recent Belle analysis of the three-body decays $\Upsilon(5S) \to [B^{(*)} \bar{B}^{(*)}]^{\mp} \pi^\pm$ gives further evidence for the existence of these states. In ths paper we analyze this decay using an effective theory of $B$ mesons interacting via strong short-range interactions. Some parameters in this theory are constrained using existing data on $\Upsilon(5 S) \to B^{(*)}\bar{B}^{(*)}$ decays, which requires the inclusion of heavy quark spin symmetry (HQSS) violating operators. We then calculate the differential distribution for $\Upsilon(5S) \to B^{(*)} \bar{B}^{(*)} \pi$ as a function of the invariant mass of the $B^{(*)} \bar{B}^{(*)}$ pair, obtaining qualitative agreement with experimental data. We also calculate angular distributions in the decay $\Upsilon(5S) \to Z_b^{(\prime)} \pi$ which are sensitive to the molecular character of the $Z_b^{(\prime)}$. 
 
\end{abstract}

\maketitle

\newpage

\section{Introduction}

The Belle collaboration recently discovered two resonances, $Z_b(10610)$ and $Z_b(10650)$ --- hereafter called $Z_b$ and $Z_b^\prime$ --- in the decays $\Upsilon(5S) \to \Upsilon(nS) \pi^+ \pi^-$ for $n$ = 1, 2, or 3 and $\Upsilon(5S) \to h_b(mP) \pi^+ \pi^-$ for $m = 1$ or 2~\cite{Collaboration:2011gja} that are the first candidates for exotic bottomonium. The experimental analysis favors the quantum numbers $I^G(J^P) = 1^+ (1^+)$   for the $Z_b$ and $Z_b^\prime$ states, which implies that the $Z_b$ and $Z_b^\prime$ couple to the meson pairs $B^*\bar{B}$ and $B^* \bar{B}^*$, respectively, in an S-wave.  Because the masses of $Z_b$ and $Z_b^\prime$  are within a few MeV of the $B^* \bar{B}$ and $B^* \bar{B}^*$ thresholds, respectively, it is likely that each of these states couples strongly to its corresponding threshold, and hence takes on a molecular character. If so, the wavefunction of  $Z_b$ ($Z_b^\prime$) at long distances is dominated by a bound-state of the $B^* \bar{B} - c.c.$ ($B^* \bar{B}^*$) though at short distances it could be more complicated, possibly resembling  a conventional bottomonium state. This scenario  is particularly likely when the conventional state's energy happens to lie very close to the threshold. If the $Z_b^{(\prime)}$ are molecular in nature, heavy quark spin symmetry (HQSS) implies there should exist as yet unseen resonances called $W_{bJ}$, where $J=0,1$, and 2~\cite{Bondar:2011ev,Voloshin:2011qa}. 

Recently, Belle \cite{Adachi:2012cx} analyzed the three-body decays $\Upsilon(5S) \to B^{(*)}\bar{B}^{(*)} \pi$ and found further evidence for the existence of the $Z_b$ and $Z_b^\prime$. Resonant structures clearly appear in the invariant mass distribution of the bottom meson-antibottom meson pair in the decays $\Upsilon(5S) \to [B^{*}\bar{B}-B\bar{B}^\ast]^\mp \pi^\pm$
and $\Upsilon(5S) \to [B^{*}\bar{B}^{*}]^\mp \pi^\pm$. Amplitudes without resonant  structure are inconsistent with the data at the 8$\sigma$ level. The experimental fits used Breit-Wigner amplitudes to analyze the spectrum and extract masses and widths of the $Z_b$ and $Z_b'$. It is well-known that for two particles that are strongly interacting in the $S$-wave due to a shallow bound state near threshold, the amplitude is  not of the Breit-Wigner form.
However, the cross sections are universal when the scattering length is large compared to the range parameters, which is expected when there is a shallow bound state or an unphysical pole in the complex plane that lies close to the threshold. In this paper, we assume this is the case for $B^{*}\bar{B} - c.c.$ and $B^*\bar{B}^*$ scattering near threshold,  and use an effective field theory (EFT) we developed in
 Ref.~\cite{Mehen:2011yh} to describe the three-body decays $\Upsilon(5S) \to B^{(*)}\bar{B}^{(*)} \pi$. 
The EFT consists of contact interactions that respect HQSS whose coefficients are tuned to provide near threshold enhancements  in $B^{*}\bar{B} - c.c.$ and $B^* \bar{B}^*$ scattering. In Ref.~\cite{Mehen:2011yh} the EFT was used to derive HQSS predictions for the binding energies, partial widths, and total widths  (some of these were first derived in Refs.~\cite{Bondar:2011ev,Voloshin:2011qa}) and also calculated rates for several two-body decay rates.  The invariant mass distributions calculated in this paper  within the same  EFT provide an interesting alternative to the Breit-Wigner parametrization, and are calculated in a systematically improvable framework based on the symmetries of QCD.  For other work treating the $Z_b$ and $Z_b^\prime$ as a hadronic molecules see Refs.~\cite{Cleven:2011gp, Cleven:2013sq,Yang:2011rp,Chen:2011zv,Zhang:2011jja,Nieves:2011vw,Sun:2011uh}, for an alternative interpretation of the $Z_b$ and $Z_b'$ as tetraquarks, see 
Refs.~\cite{Ali:2011ug,Ali:2011vy,Cui:2011fj,Guo:2011gu}

In the next section of this paper, we analyze the decays  $\Upsilon(5S) \to B^{(*)} \bar{B}^{(*)}$ and  $\Upsilon(5S) \to B^{(*)} \bar{B}^{(*)} \pi$. We determine some of the couplings in our EFT by fixing parameters using available data on the decays $\Upsilon(5S) \to B^{(*)} \bar{B}^{(*)}$. However, to obtain quantitive agreement with observed branching ratios  requires that we include HQSS violating operators in addition to the terms respecting HQSS.  We also analyze $\Upsilon(5S) \to B^{(*)} \bar{B}^{(*)} \pi$. The $B$ and $\bar{B}$ mesons are strongly interacting in the $B^* \bar{B} - c.c.$ and $B^* \bar{B}^*$ channels, so in these channels tree-level graphs must be augmented by loop diagrams which include the leading contact interaction to all orders. These loops give the structure in the amplitude to obtain the $Z_b$ and $Z_b^{\prime}$ resonances. The theory can accomodate  the relatively large branching ratio for $\Upsilon(5S) \to B^* \bar{B} \pi, \bar{B} B^* \pi$ observed experimentally in Ref.~\cite{Drutskoy:2010an}. Previous theoretical analyses of $\Upsilon(5S) \to B^{(*)} \bar{B}^{(*)}  \pi$~ failed to predict this large branching ratio~\cite{Lellouch:1992bq,Simonov:2008cr}. 

Once the relevant coupling constants are constrained using two-body and three-body decays of the $\Upsilon(5S)$, we then consider angular distributions in the decays $\Upsilon(5S) \to Z_b^{(\prime)} \pi$ in the following section.
In $e^+e^- \to \Upsilon(5S)$ the $\Upsilon(5S)$ is produced with polarization transverse to the beam. Therefore, the decay rate is not isotropic and the decay rate for $\Upsilon(5S) \to Z_b^{(\prime)} \pi$  depends on the angle the pion makes with the beam axis, $\theta$, as 
\bea\label{eqn:rho}
\frac{d\sigma}{d \cos\theta} \propto 1+ \rho_{Z^{(\prime)}} \, \cos^2 \theta \, ,
\eea
where $-1 \leq  \rho_{Z^{(\prime)}} \leq 1$. In the heavy quark limit, HQSS   predicts that the rates  $\Gamma[\Upsilon(5S) \to Z_b \pi]$ and $\Gamma[\Upsilon(5S) \to Z_b^{\prime} \pi]$ are equal and that $ \rho_{Z^{(\prime)}} = 0$. More interesting is the pattern of HQSS violation. 
 In this case, the leading HQSS breaking corrections to short-distance contributions to the decays change the relative rates but still yield $ \rho_{Z^{(\prime)}}=0$. However, long-distance contributions in which the pion couples to one of the constituent $B$ mesons, can yield nonvanishing but small $\rho_{Z^{(\prime)}}$. Thus, measuring non-vanishing $\rho_{Z^{(\prime)}}$ with a value consistent with our calculations is evidence for the molecular character of these states. However, the values of $\rho_{Z^{(\prime)}}$ we obtain from the fits in this paper turn out to be very small, with $\rho_Z$ ranging from $0.001$ to $0.03$ and $\rho_{Z'} =-0.02$, and will be difficult to observe.

Following this section are our conclusions.

\section{$\Upsilon(5S)$ Decays to $B^{(*)} \bar{B}^{(*)}$ and $B^{(*)} \bar{B}^{(*)} \pi$}

The relevant terms in the HH$\chi$PT Lagrangian are
\begin{align}\label{eqn:HHchiPT_Lagrangian}
\calL_{\rm HH\chi PT} &= \tr(H_a^\dagger i\d_0 H_a) + \tfrac{1}{4}\Delta \tr(H^\dagger_a \sigma_i H_a \sigma^i) + \tr(\bar H_a^\dagger i\d_0 \bar H_a) + \tfrac{1}{4}\Delta \tr(\bar H^\dagger_a \sigma_i \bar H_a \sigma^i)\\
&+ g\,\tr(\bar H_a \bar H^\dagger_b \bsigma)\cdot\bfA_{ab} - g\,\tr(H_a^\dagger H_b \bsigma)\cdot\bfA_{ab} \nonumber\\
&+ \tfrac{1}{2} [g_{\Upsilon}\,\tr(\Upsilon \bar H^\dagger_a \bsigma\cdot i\lrbd H_a^\dagger) + g_{\Upsilon\pi}\tr(\Upsilon \bar H_a^\dagger H_b^\dagger)A^0_{ab}] \nonumber \\
&+ \tfrac{1}{4}g_1 \tr [ (\Upsilon \sigma^i +\sigma^i \Upsilon)\bar{H}_a^\dagger  i\lrbd_i H_a^\dagger ] 
+ \tfrac{1}{4}g_2 \tr [(\sigma^i  \Upsilon \sigma^j +\sigma^j \Upsilon \sigma^i)\bar{H}_a^\dagger \sigma^i   i\lrbd_j H_a^\dagger ] \nonumber \\
&+ \tfrac{1}{4} g_{\Upsilon \pi}^\prime \tr[(\Upsilon \sigma^i +\sigma^i \Upsilon) \bar{H}_a^\dagger \sigma^i H_a^\dagger] A^0 + {\rm h.\,c.} \nn \,,
\end{align}
which are the given in \cite{Mehen:2011yh} except for the last three terms which are added to break HQSS in the Lagrangian. In Eq.~(\ref{eqn:HHchiPT_Lagrangian}), the fields for $B^{(*)}$ and $\bar B^{(*)}$ mesons we use the $2\times2$ matrix notation described in Ref.~\cite{Hu:2005gf}, where $H_a = \bfB^\ast_a\cdot\bsigma + B_a{\bf1}$,
$ \bfB^\ast_a$ and $B_a$ are vectors and pseudoscalars, respectively, and 
 $a$ is an antifundamental index describing the flavor of the light antiquark bound to the bottom quark.  Therefore, $H_1$ contains the $B^-$ and $B^{*-}$, $H_2$ has the $\bar B^0$ 
 and $\bar{B}^{*0}$, while $\bar H_1$ and $\bar H_2$ contain their respective antiparticles. The $\Upsilon(5S)$ has $I^G(J^{PC}) = 0^-(1^{--})$ and is paired under HQSS with the pseudoscalar $\eta_b(5S)$.  These appear in the  $2\times2$ matrix field  $\Upsilon = {\bf\Upsilon}(5S)\cdot\bsigma + \eta_b(5S){\bf 1}$.

\begin{figure}[t]
\begin{center}
\includegraphics[width=6.75in,trim=3cm 18cm 3cm 4cm,clip]{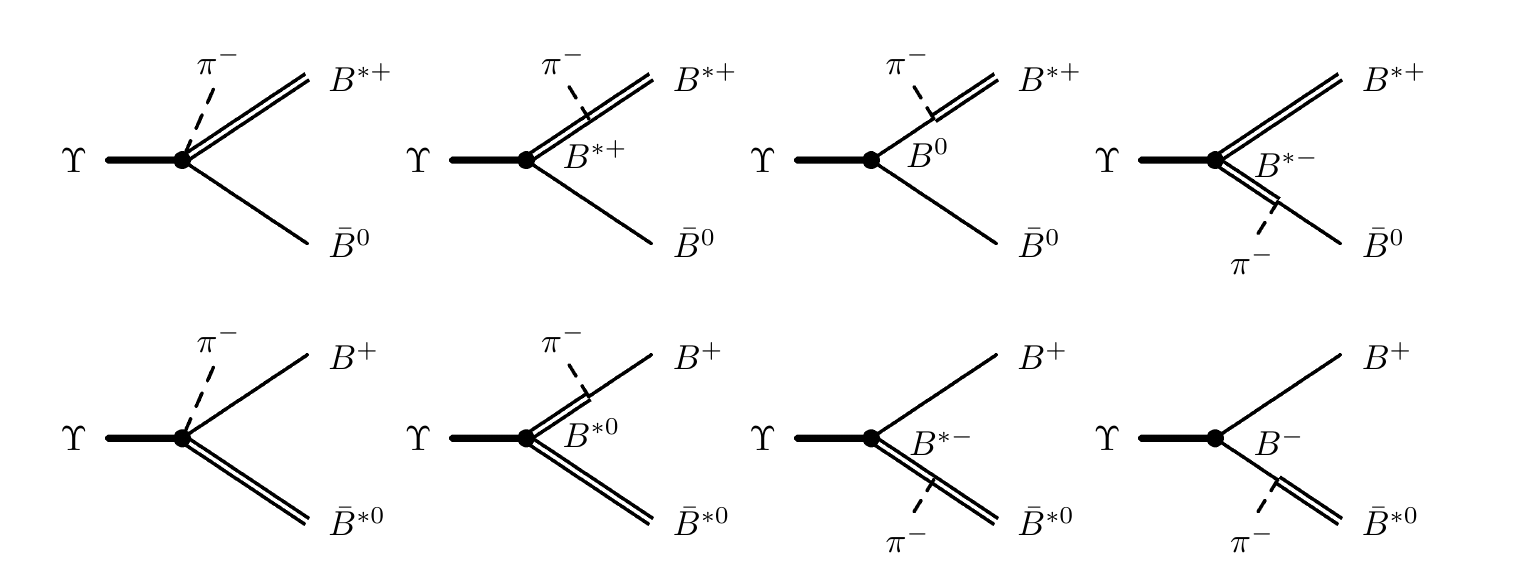}
\caption{The eight diagrams contributing to $\Upsilon(5S) \to B^+ \bar{B}^{*0}, B^{*+} \bar{B}^{0}$ and  and $\Upsilon (5S) \to  Z_{b}^+ \pi^-$. }
\label{fig:UpsilonZb}
\end{center}
\end{figure}

The first line of Eq.~(\ref{eqn:HHchiPT_Lagrangian}) consists of the kinetic terms for $B^{(*)}$ and $\bar{B}^{(*)}$ and the terms that give rise to the hyperfine splittings. The second line has the axial couplings to pions. The coupling constant $g$ is known to be $0.6 \pm 0.1$ from a tree-level analysis of strong $D^*$ meson decays. The third line has the couplings involving the  $\Upsilon(5S)$. The term with coupling constant
$g_{\Upsilon}$ couples the $\Upsilon(5S)$ to the heavy mesons. The term with coupling constant  $g_{\Upsilon \pi}$ is a four-field contact interaction that couples the $\Upsilon(5S)$, the $B^{(*)}$ and $\bar B^{(*)}$ mesons, and the pion. Both interactions contribute to the decays at leading order. This is because the tree-level diagram with the contact interaction, e.g., the figures on the left in Fig.~\ref{fig:UpsilonZb}, have one time derivative  which contributes a factor of $E_\pi$, where $E_\pi$ is the pion energy,  to the amplitude. Tree level diagrams with the interaction proportional to $g_{\Upsilon}$,  e.g., the remaining diagrams in Fig.~\ref{fig:UpsilonZb}, have derivatives at both vertices giving a factor of $p_\pi^2$, where $p_\pi$ is the pion momentum,   but also a factor  $\propto E_\pi^{-1}$ due to the energy dependence of the meson propagator. Thus both diagrams scale as $Q$ where 
$Q \sim p_\pi \sim E_\pi$.  The second to last line of Eq.~(\ref{eqn:HHchiPT_Lagrangian}) contain the HQSS violating couplings of the $\Upsilon(5S)$ to the heavy mesons and the last line contains the HQSS violating couplings of the $\Upsilon(5S)$ to heavy mesons and pions. One can check that these are the only operators of this dimension that are consistent with all symmetries other than HQSS (see Ref.~\cite{Fleming:2008yn} for a complete listing of symmetries and field transformations).

From our Lagrangian we calculate the following rates for the two-body decays of the $\Upsilon(5S)$:
\bea
\Gamma[\Upsilon(5S)\to B\bar{B}] &=& \frac{p_B^3}{6\pi} \frac{m_B^2}{m_{\Upsilon(5S)}}(g_\Upsilon + g_1 +3 g_2)^2  \\
\Gamma[\Upsilon(5S)\to B^*\bar{B}] = \Gamma[\Upsilon(5S)\to B \bar{B}^*] &=& \frac{p_B^3}{3\pi} \frac{m_B m_{B^*}}{m_{\Upsilon(5S)}}(g_\Upsilon - 2 g_2)^2 \nn\\
\Gamma[\Upsilon(5S)\to B^*\bar{B}^*] &=& \frac{p_B^3}{6\pi} \frac{m_{B^*}^2}{m_{\Upsilon(5S)}}\left(\tfrac{20}{3} g_\Upsilon^2 +3 (
\tfrac{1}{3}g_\Upsilon -g_1+ g_2)^2 \right)\nn \, .
\eea
Here $p_B$ is the momentum of the $B^{(*)}$ meson in the decay. In the HQSS limit one finds
$\Gamma[\Upsilon(5S)\to B\bar{B}]:\Gamma[\Upsilon(5S)\to B\bar{B}^* + \bar{B} B^*]:\Gamma[\Upsilon(5S) \to B^* \bar{B}^*]::1:4:7$.
Upon including the kinematic factors of $p_B^3$ appropriate for each decay, this becomes $1:3.2:4.3$. The central values of the experimental branching ratios
are in the ratio $1: 2.5\pm0.5:6.9\pm1.4$, so violations of HQSS are important for these observables.
We fit the parameters $g_\Upsilon$, $g_1$, and $g_2$ to the product of branching fractions and total width for the $\Upsilon(5S)$ given in the PDG~\cite{Beringer:1900zz} and find
\bea\label{coupling}
g_{\Upsilon} = 0.112 \, {\rm GeV}^{-3/2}, \quad g_1 = - 0.048 \, {\rm GeV}^{-3/2}, \quad g_2 = 0.012 \,  {\rm GeV}^{-3/2} \, .
\eea
The uncertainty in the total width of the $\Upsilon(5S)$ is $51\%$, the uncertainties in the branching ratios are significantly smaller ($< 18 \%$).
We conclude that uncertainties in the coupling constants in Eq.~(\ref{coupling}) are of order 25\%. We will use the values in Eq.~(\ref{coupling}) in our analysis below. Since the couplings of the operators with coefficients $g_1$ and $g_2$ violate HQSS, we expect these constants to be suppressed by 
$\Lambda_{\rm QCD}/m_B \sim 0.1-0.2$. The coupling constant $g_1$ exceeds this by a factor of $\sim 2-4$, while $g_2$ is in line with our expectations.

The decays $\Upsilon(5S) \to B^{(*)} \bar{B}^{(*)}$ were recently analyzed in 
Ref.~\cite{Meng:2007tk} which uses a relativistic formalism whose non-relativistic limit is equivalent to our EFT. Corrections to the  nonrelativistic approximation  should be small since in the two-body decays the velocity of the $B$-mesons is $v =0.22-0.24$ and corrections typically scale as $v^2 =0.05-0.06$.  
In the HQSS limit, $g_{\Upsilon B B} = g_{\Upsilon B^* B}=g_{\Upsilon B^* B^*}$, and so the authors of Ref.~\cite{Meng:2007tk} incorporate HQSS violation in the decays  $\Upsilon (5S) \to B^{(*)} \bar{B}^{(*)}$ by using the Feynman rules obtained from the leading HQSS operator, but letting the coupling constants $g_{\Upsilon B B}$, $g_{\Upsilon B^* B}$ and $g_{\Upsilon B^* B^*}$ differ for each decay. In  our analysis of $\Upsilon(5S) \to B\bar{B}$ and  $\Upsilon(5S) \to B^*\bar{B} - c.c.$, the effect of the leading HQSS operators  is simply to change the coupling constants: $g_\Upsilon \to g_\Upsilon +g_1 +3 g_2$
for $\Upsilon(5S) \to B \bar{B}$ and $g_\Upsilon \to g_\Upsilon - 2 g_2$ for $\Upsilon(5S) \to B^* \bar{B}, B \bar{B}^*$. However, in the case of $\Upsilon(5S) \to B^*\bar{B}^*$, HQSS violation leads to new structures in the matrix element. The tree-level amplitude is 
\bea
\epsilon^{B^*}_i \epsilon^{\bar{B}^*}_j  \epsilon^\Upsilon_k \left[
g_\Upsilon\left( -(p^i_B - p^i_{\bar{B}}) \delta^{jk} - (p^j_B - p^j_{\bar{B}}) \delta^{ik} 
+ (p^k_B - p^k_{\bar{B}}) \delta^{ij} \right) +(g_2-g_1)  (p^k_B - p^k_{\bar{B}}) \delta^{ij} \right] \, ,
\eea
where $p_{\bar{B}}$ is the momentum of the $\bar{B}^*$, and $\epsilon^{B^*}$, $\epsilon^{\bar{B}^*}$, and $ \epsilon^\Upsilon$ are the polarization vectors of the $B^*$, $\bar{B}^*$ and $\Upsilon$, respectively. The tensor structure of the operator changes when the coefficients $g_1$ and $g_2$ are nonzero so simply changing the value of $g_\Upsilon$ in this amplitude does not properly account for the leading HQSS violating effects.

Next we turn to the calculation of $\Upsilon(5S) \to B^{(*)} \bar{B}^{(*)}\pi$. For the decay $\Upsilon(5S) \to B \bar{B} \pi$ there are two diagrams (not shown) in which $\Upsilon(5S) \to B^* \bar{B}$ (or $B \bar{B}^*$) is followed by $B^* \to B \pi$ (or   $\bar{B}^* \to \bar{B} \pi$). There is no contribution to the decay from tree-level contact diagrams. 
Furthermore, there are no strong interactions in the $B \bar{B}$ channel as the contact interactions that are nonperturbative exist only in the $B^* \bar{B} - c.c.$ and $B^* \bar{B}^*$ channels.
The expression we find for the three-body decay rate is \cite{Lellouch:1992bq}
\bea
\frac{d^2\Gamma[\Upsilon(5S) \to B^+ \bar{B}^0 \pi^-]}{dE_B dE_{\bar{B}} }=\frac{g^2(g_\Upsilon-2 g_2)^2 m_B m_{\bar{B} }}{12 \pi^3 f^2}
\frac{p_B^2 p_{\bar{B}}^2 -(\vec p_B \cdot \vec p_{\bar{B}})^2}
{(E_\pi-\Delta)^2} \, .
\eea
Here $f = 132$ MeV is the pion decay constant, $E_\pi$ is the energy of the pion, and $\Delta = 42$ MeV is the hyperfine splitting of the $B$ mesons. The rate for final states with neutral pions is 1/2 the rate for charged pions. Integrating over phase space and summing over the final states $B^+ \bar{B}^0 \pi^-$,
$B^0 B^- \pi^+$, $B^0 \bar{B}^0 \pi^0$ and $B^+ B^- \pi^0$, we find $\Gamma[\Upsilon(5S) \to B \bar{B}\pi] = 0.03$ MeV. Using the PDG expression for the total width of the 
$\Upsilon(5S)$ yields a branching fraction of $5.5^{+5.7}_{-1.8}\,\times 10^{-4}$, which is roughly an order of magnitude below the limit of $4.0 \,\times10^{-3}$ obtained in Ref.~\cite{Adachi:2012cx} .

\begin{figure}[t]
\begin{center}
\includegraphics[width=6.75in,trim=2.2cm 21cm 2.2cm 3.4cm,clip]{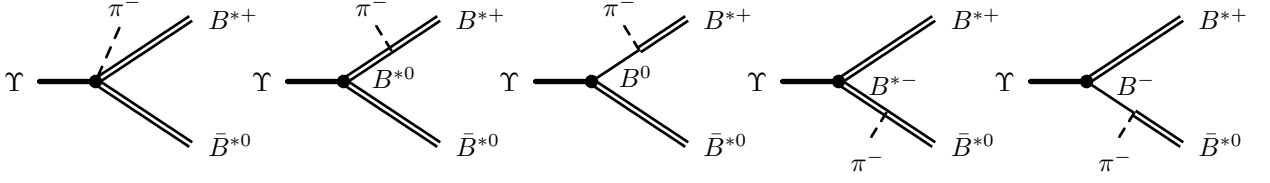}
\caption{The five diagrams contributing to $\Upsilon(5S) \to B^{*+} \bar{B}^{*0}$ and $\Upsilon (5S) \to  Z_{b}^{\prime +} \pi^-$. }
\label{eqn:UpsilonZbpPi}
\end{center}
\end{figure}

The tree-level diagrams for $\Upsilon(5S) \to B^{*+} \bar{B}^{0}\pi^-,B^{+} \bar{B}^{*0}\pi^-$ 
 are shown in Fig.~\ref{fig:UpsilonZb} and the diagrams for 
$\Upsilon(5S) \to B^{*+} \bar{B}^{*0}\pi^-$ are shown in Fig.~\ref{eqn:UpsilonZbpPi}. 
The corresponding tree-level amplitude for $\Upsilon(5S) \to B^{*+} \bar{B}^0 \pi^-$  is given by:
\bea
i {\cal M}^{\rm tree}[\Upsilon(5S) \to B^{*+} \bar{B}^0 \pi^-] &=& \\
&&\hspace{-0.75 in}\epsilon_{\Upsilon}^i \epsilon_{B^*}^{j *}  \nn  \left( A^{\rm tree}_1 \, \delta^{ij} + A^{\rm tree}_2 \, p_B^i p_B^j 
+ A^{\rm tree}_3\, p_{\bar B}^i p_{\bar B}^j  + A^{\rm tree}_4 \, p_{\bar B}^i p_B^j + A^{\rm tree}_5 \, p_B^i p_{\bar B}^j  \right)  ,
\eea
where the functions $A^{\rm tree}_i$ are 
\bea\label{atree}
A^{\rm tree}_1 &=& -(g_{\Upsilon \pi} + g_{\Upsilon \pi}^\prime) \frac{E_\pi}{f_\pi} +\frac{2 g(g_\Upsilon -2 g_2)}{ f_\pi \, E_\pi} \vec{p}_\pi \cdot \vec{p}_{\bar B} +
\frac{2 g g_\Upsilon}{f\,(E_\pi - \Delta)} \,\vec{p}_\pi \cdot \vec{p}_B \\
A^{\rm tree}_2 &=& \frac{2 g (g_2-g_1)}{f_\pi \,(E_\pi -\Delta)}  \nn \\
A^{\rm tree}_3 &=&   \frac{2 g(g_\Upsilon - 2g_2)}{f_\pi \,E_\pi} -  \frac{2 g(g_\Upsilon + g_1 +3g_2)}{f_\pi(E_\pi +\Delta)}\nn \\
A^{\rm tree}_4 &=& -\frac{2 g(g_\Upsilon +g_1+3 g_2)}{f_\pi \,(E_\pi + \Delta)} - \frac{2 g g_\Upsilon}{f_\pi \,(E_\pi -\Delta)}  \nn \\
A^{\rm tree}_5 &=& \frac{2 g(g_\Upsilon -2 g_2)}{f_\pi \,E_\pi} + \frac{2g(g_\Upsilon +g_2 -g_1)}{f_\pi \,(E_\pi -\Delta)} \nn \, .
\eea
The tree-level amplitude for $\Upsilon(5S) \to B^+\bar{B}^{*0} \pi^-$ differs only by an overall sign and the replacement $p_B \leftrightarrow p_{\bar{B}}$.

The tree-level amplitude for $\Upsilon(5S) \to B^{*+} \bar{B}^{*0} \pi^-$ is 
\bea
i{\cal M}^{\rm tree}[\Upsilon(5S) \to B^{*+} \bar{B}^{*0} \pi^-] &=& \\
&& \hspace{-1.75 in} -i  \epsilon_{\Upsilon}^i \epsilon_{B^*}^{j *}  \epsilon_{\bar{B}^*}^{k *} \big(B^{\rm tree}_1 \,\epsilon^{ijk}  + B^{\rm tree}_2 \, [\epsilon^{ijl}  p_B^l  (p_B + p_{\bar B})^k -\epsilon^{ikl}  p_{\bar B}^l  (p_B + p_{\bar B})^j ] \nn \\
&&\hspace{-1.75 in}+ B^{\rm tree}_3 \, \epsilon^{kjm} (p_B + p_{\bar B})^m (p_B + p_{\bar B})^i  \nn \\
&& \hspace{-1.75 in}+ B^{\rm tree}_4 \, \left[ \delta^{ik} \epsilon^{jlm} p_{\bar B}^l \, p_B^m   -\delta^{ij} \epsilon^{klm} p_B^l \,p_{\bar B}^m + \epsilon^{jim} (p_B+ p_{\bar B})^m p_{\bar B}^k - \epsilon^{kim} (p_B+ p_{\bar B})^m p_B^j \right] \big) \nn\, ,
\eea
where 
 \bea 
B^{\rm tree}_1 &=& \frac{(g_{\Upsilon \pi} - g_{\Upsilon \pi}^\prime)}{f_\pi} E_\pi \\
 B^{\rm tree}_2 &=&  \frac{2 g(g_\Upsilon - 2 g_2)}{f_\pi(E_\pi + \Delta)} \nn \\
 B^{\rm tree}_3 &=& -\frac{2 g(g_\Upsilon +g_2 -g_1)}{f_\pi \,E_\pi} \nn \\
 B^{\rm tree}_4 &=& - \frac{2 g g_\Upsilon}{f_\pi\,E_\pi}  \, .\nn
 \eea
It is helpful to separate this amplitude into pieces that are symmetric and  antisymmetric under $\epsilon_{B^*}^{j *} \leftrightarrow  \epsilon_{\bar{B}^*}^{k *}$. The antisymmetric piece of this amplitude contributes to final states with  $B^*$ and $\bar{B}^*$ in a $S=1$ spin state.
The $Z_b$ and $Z_b^\prime$ can only appear in this channel, so only this channel will be modified by final-state rescattering effects. If we make the replacement $\epsilon_{B^*}^{j *} \epsilon_{\bar{B}^*}^{k *}
\rightarrow \frac{i}{\sqrt{2}}\epsilon^{jka}\epsilon_{B^*\bar{B}^*}^{a*}$, we find
\bea
i{\cal M}^{\rm tree}[\Upsilon(5S) \to (B^{*+} \bar{B}^{*0})_{S=1}  \pi^-] =  \frac{\epsilon_{\Upsilon}^i \epsilon_{B\bar{B}}^{a*}}{\sqrt{2}}
\big( B^{\rm tree}_5  \,\delta^{ia} + B^{\rm tree}_6 \, p_\pi^i   p_\pi^a \big) \,  ,
\eea
 where $B^{\rm tree}_5= 2 B^{\rm tree}_1 + B^{\rm tree}_2 \, p_\pi^2  -B^{\rm tree}_4 \, p_\pi^2$, $B^{\rm tree}_6 = B^{\rm tree}_4 -B^{\rm tree}_2 -2 B^{\rm tree}_3$ and $\epsilon_{B^*\bar{B}^*}^{a*}$ is a polarization vector for the combined $B^* \bar{B}^*$ system. n the CM frame $\vec{p}_\pi = - \vec{p}_B - \vec{p}_{\bar B}$.
 
We can square the $S=1$ and $S\neq 1$ pieces of the amplitude separately. For $S=1$ the result is 
\bea
\frac{1}{3}\sum |{\cal M}^{\rm tree}[\Upsilon(5S)\!&\to&\!(B^{*+} \bar{B}^{*0})_{S=1}  \pi^-]|^2\\
&&= \tfrac{1}{6}\big(3 |B^{\rm tree}_5|^2 + 2\, {\rm Re}[ (B_5^{\rm tree})^* B^{\rm tree}_6]\, p_\pi^2 +|B^{\rm tree}_6|^2 (p_\pi^2)^2\big) \, .  \nn
\eea
Note the $B^{\rm tree}_i$ are real at tree level but $B_5^{\rm tree}$ and $B_6^{\rm tree}$ will be replaced by complex numbers when we include higher order corrections, so we start to treat them as complex numbers even in this formula.
For $S\neq1$ we find
\bea 
\frac{1}{3}\sum|{\cal M}^{\rm tree}[\Upsilon(5S)\!&\to&\! (B^{*+} \bar{B}^{*0})_{S\neq1}  \pi^-]|^2 = \\
&& \tfrac{1}{12}\big[ (B^{\rm tree}_2)^2 (6 \, p_\pi^2 (\vec{p}_B-\vec{p}_{\bar B})^2 -2 \, (p_B^2 - p_{\bar B}^2)^2) \nn \\
&+& (B^{\rm tree}_4)^2 \big( 56\, (p_B^2 \,p^2_{\bar B}- (\vec{p}_B \cdot \vec{p}_{\bar B})^2) + 4 \, p_\pi^2 \,(\vec{p}_B-\vec{p}_{\bar B})^2 \big) \nn \\
&+& 2  B_2^{\rm tree} B^{\rm tree}_4 \big( 8\, ( (\vec{p}_B \cdot \vec{p}_{\bar B})^2 -p_B^2 p_{\bar B}^2) + 4\, (p_B^2 - p_{\bar B}^2)^2 \big) \big] \,.\nn
\eea

\begin{figure}[t]
\begin{center}
\includegraphics[width=6.75in,trim=2.2cm 21cm 2.2cm 3.4cm,clip]{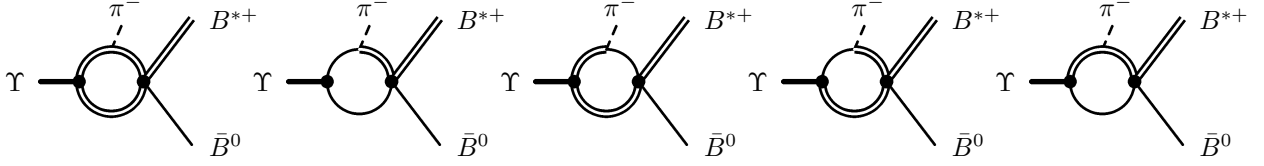}
\caption{Five one-loop diagrams contributing to $\Upsilon(5S) \to B^{*+} \bar{B}^{0} \pi^-$. }
\label{oneloop}
\end{center}
\end{figure}

Because the $B$ mesons in the final state are strongly interacting we have to consider diagrams with an arbitrary number of insertions of the leading order contact interactions.   We only consider diagrams where $B^{(*)}\bar{B}^{(*)}$ 
rescatter  after the emission of a pion. Before the pion emission, the $B^{(*)}\bar{B}^{(*)}$ pair has an invariant mass equal to $M_{\Upsilon(5 S)}$, so are far from the threshold and hence resumming contact interaction is unimportant. 
The effect of resumming the contact interactions before the pion is emitted yields a set of diagrams that is identical to what is obtained if  the contact interactions are resummed in the decays $\Upsilon(5S) \to B^{(*)} \bar{B}^{(*)}$.  The effect of these diagrams in both cases can be absorbed into the definition of the couplings $g_\Upsilon$, $g_1$, and $g_2$. On the other hand, final state interactions will depend on the invariant mass of the $B^{(*)}$ and $\bar{B}^{(*)}$ mesons in the final state and will give rise to the resonant structure in the amplitudes.
The one-loop  diagrams  for $\Upsilon(5S)\to B^* \bar{B} \pi$ with one contact interaction after the emission of the pion are shown are shown in Fig.~\ref{oneloop}. The diagrams $\Upsilon(5S)\to B^{*} \bar{B}^* \pi$ are identical except the final state 
$\bar{B}$ is replaced with a $\bar{B}^*$.
For the one-loop diagrams for $\Upsilon(5S)\to B^{*+} \bar{B}^0 \pi^-$ we find 
\bea
i{\cal M}^{\rm one-loop}[\Upsilon(5S)\to B^{*+} \bar{B}^0 \pi^-]&=& \frac{g\,  \bar{m}_B^{3/2}}{8\pi f }\epsilon_{\Upsilon}^i \epsilon_{B^*}^{j *}
\big( C_1 \, p_\pi^2  \delta^{ij} +  C_2 \, p_\pi^i  p_\pi^j \Big) \, ,
\eea
where $\bar{m}_B =(3 \,m_{B^*}+m_B)/4$ is the spin-averaged $B$-meson mass,  and $C_1$ and $C_2 $ are given by 
\bea
C_1 &=&  \frac{C_- (g_\Upsilon -2 g_2)}{\sqrt{b_{BB^*}}}\,F\left(\frac{E_\pi+\Delta}{b_{BB^*}}\right)
+ \frac{C_- g_\Upsilon }{\sqrt{b_{B^*B^*}}}\,F\left(\frac{E_\pi}{b_{B^*B^*}}\right)\\
&& - \frac{C_+ (g_\Upsilon-2 g_2) }{\sqrt{b_{BB^*}}}\,F\left(\frac{E_\pi}{b_{BB^*}}\right) 
- \frac{C_+ g_\Upsilon }{\sqrt{b_{B^*B^*}}}\,F\left(\frac{E_\pi-\Delta}{b_{B^*B^*}}\right)\, ,\nn
\\
\nn\\
C_2 &=& - \frac{C_+ (g_\Upsilon + g_1+ 3 g_2 )}{\sqrt{b_{BB}}}\,F\left(\frac{E_\pi+\Delta}{b_{BB}}\right)
-  \frac{C_- (g_\Upsilon-2 g_2) }{\sqrt{b_{BB^*}}}\,F\left(\frac{E_\pi+\Delta}{b_{BB^*}}\right) \nn \\
&&
+  \frac{C_- (g_\Upsilon +2  g_2 - 2 g_1) }{\sqrt{b_{B^*B^*}}}\,F\left(\frac{E_\pi}{b_{B^*B^*}}\right) 
+ \frac{C_+ (g_\Upsilon-2 g_2) }{\sqrt{b_{BB^*}}}\,F\left(\frac{E_\pi}{b_{BB^*}}\right) \nn\\
&& + \frac{C_+ (g_2-g_1) }{\sqrt{b_{B^*B^*}}}\,F\left(\frac{E_\pi-\Delta}{b_{B^*B^*}}\right)\, . \nn
\eea
Here $b_{B^{(*)}B^{(*)}}= m_{\Upsilon(5S)} - m_{B^{(*)}}-m_{B^{(*)}}$ and the function $F(x)$ is given by
\bea
F(x) &=& \int_0^1 dy \frac{y}{\sqrt{-1+ x y - i \epsilon}}  \\
&=& i\left( \frac{4 - (4+2x) \sqrt{1-x}}{3 x^2}\right) \qquad (x<1) \nn \\
&=&  \frac{(4+2x) \sqrt{x-1}+ i 4}{3 x^2}  \qquad \qquad (x>1) \, .\nn 
\eea
In evaluating the loop integrals we drop terms suppressed by $p_\pi^2/(\bar{m}_B b_{B^{(*)}B^{(*)}}) \approx 0.05$.
Here $C_\pm = C_{10}\, \pm \,C_{11}$, where $C_{10}$ and $C_{11}$ were defined in Ref.~\cite{Mehen:2011yh}. 
The loop diagrams for $\Upsilon(5S) \to B^* \bar{B}^* \pi$ only contribute to $S=1$ final states. Therefore, we can make the 
replacement $\epsilon_{B^*}^{j *} \epsilon_{\bar{B}^*}^{k *} \rightarrow \frac{i}{\sqrt{2}}\epsilon^{jka}\epsilon_{B\bar{B}}^{a*}$ in computing 
this amplitude.  Upon making  this replacement, we find that $i{\cal M}^{\rm one-loop}[\Upsilon(5S) \to (B^{*+} \bar{B}^{*0})_{S=1} \pi^-] = i \sqrt{2} {\cal M}^{\rm one-loop}[\Upsilon(5S) \to B^{*+} \bar{B}^0 \pi^- ]$ after replacing 
$\epsilon_{B}^{j*}$ with $\epsilon_{B^*\bar{B}^*}^{j*}$ and interchanging $C_+ \leftrightarrow C_-$. 

Next we consider the effect of final state interactions on the amplitudes. The tree-level diagrams need their outgoing $B^{(*)}\bar{B}^{(*)}$ mesons dressed with strong contact interactions.
These diagrams dress the tree-level contact interactions proportional to 
$g_{\Upsilon \pi} \pm g_{\Upsilon \pi}^\prime$and the one-loop diagrams. 
 The diagrams in which one adds contact interactions in the final state 
to tree-diagrams with virtual $B^{(*)}$ mesons are the loop diagrams and their dressing.

Let 
\bea 
i{\cal M}= \left( \begin{array}{c} i {\cal M}_{B^*B^*} \\  i{\cal M}_{B B^*} \end{array}\right) \, ,
\eea
be a vector constructed from the amplitudes for final states with $B^* \bar{B}^*$ or $B^* \bar{B}^*$. Let $C$ represent the matrix of contact interactions \cite{Mehen:2011yh}
\bea\label{Cmatrix}
C = \left( \begin{array}{cc} C_+ & C_- \\  C_- & C_+ \end{array}\right) \, ,
\eea
and let $\Sigma_Z$ be 
\bea
\Sigma_Z = \left( \begin{array}{cc} \Sigma_{B^* B^*}(E) &0\\ 0 & \Sigma_{B B^*}(E)   \end{array}\right) \, ,
\eea
where the functions $\Sigma_{B^* B^*}(E) $ and $\Sigma_{B B^*}(E)$ are defined in Ref.~\cite{Mehen:2011yh}.
Then the dressing of these amplitudes with contact interactions leads to an amplitude  given by the infinite matrix series:
\bea\label{dressing}
i {\cal M}^{\rm dressed} &=&\left( 1 -C\, \Sigma_Z  + C\, \Sigma_Z \, C \, \Sigma_Z  + \ldots  \right)\, i{\cal M} \\
&=& (1 + T_Z \Sigma_Z ) \, i {\cal M} \nn \\
&=& - T_Z \, C^{-1} \,i {\cal M}   \nn \, .
\eea
Here $T_Z$ is the $T$-matrix calculated in Ref.~\cite{Mehen:2011yh}:
\bea
T_Z= \left( \begin{array}{cc} T_{Z'Z'}(E) & T_{Z'Z}(E) \\  T_{ZZ'}(E) & T_{ZZ}(E) \end{array}\right) \, .
\eea
where 
\bea\label{Tmatrix}
T_{Z^\prime Z^\prime}(E) &=&\frac{4 \pi}{\bar{m}_B} \frac{-\gamma_+ +\sqrt{\bar{m}_B(\Delta-E)-i\epsilon}}{(\gamma_+ -\sqrt{\bar{m}_B(\Delta-E)-i\epsilon})(\gamma^{\prime}_+ -\sqrt{\bar{m}_B(2\Delta-E)-i\epsilon}) -\gamma_-^2 } \\
T_{Z^\prime Z}(E) &=& T_{Z Z^\prime}(E) =\frac{4 \pi}{\bar{m}_B} \frac{\gamma_- }{(\gamma_+ -\sqrt{\bar{m}_B(\Delta-E)-i\epsilon})(\gamma^{\prime}_+ -\sqrt{\bar{m}_B(2\Delta-E)-i\epsilon}) -\gamma_-^2 } \nn
\\
T_{ZZ}(E) &=&\frac{4 \pi}{\bar{m}_B}
\frac{-\gamma^{\prime}_+ +\sqrt{\bar{m}_B(2\Delta-E)-i\epsilon}}{(\gamma_+ -\sqrt{\bar{m}_B(\Delta-E)-i\epsilon})(\gamma^{\prime}_+ -\sqrt{\bar{m}_B(2\Delta-E)-i\epsilon}) -\gamma_-^2 } \, , \nn
\eea
In this formula, $\gamma_+^{(\prime)}$ and $\gamma_-$ determine the location of the $Z_b$ and $Z_b^\prime$ relative to their thresholds. 
These parameters can be chosen to be complex, giving the molecular states a finite width. 
In the HQSS limit $\gamma_+ = \gamma_+^\prime$~\cite{Mehen:2011yh}.  Here we have allowed for the possibility of HQSS violation in the contact interaction.
While it is in principle possible to repeat the analysis of Ref.~\cite{Mehen:2011yh} including HQSS  violating contact interactions, it is easy to see that the most general $2\times2$ matrix  that can replace 
$C$ in Eq.~(\ref{Cmatrix}) will be symmetric and have different coefficients in the two terms along the diagonal. Then repeating the analysis of Ref.~\cite{Mehen:2011yh} one obtains the $T$-matrices in Eq.~(\ref{Tmatrix}) 
with $\gamma_+ \neq \gamma_+^\prime$. Later in the paper we will choose $\gamma_+^{(\prime)}$ and $\gamma_-$  so that the poles in $T_Z$ are located at the complex energies determined by other experimental or theoretical analyses.

The loop amplitudes can be written as
\bea\label{loop}
i {\cal M}^{\rm 1-loop} &=&  \left( \begin{array}{c} i {\cal M}^{\rm 1-loop}_{B^*B^*} \\  i{\cal M}^{\rm 1-loop}_{BB^*} \end{array}\right) \\
&=&
 \left( \begin{array}{cc} C_+ & C_- \\  C_- & C_+ \end{array}\right) 
 \left( \begin{array}{c} L_{Z'}^1(E_\pi) \, p_\pi \cdot \epsilon_\Upsilon \, p_\pi \cdot \epsilon_{Z'}  + L^2_{Z'}(E_\pi) \, p_\pi^2 \,\epsilon_\Upsilon  \cdot \epsilon_{Z'}
 \\ L_Z^1(E_\pi) \, p_\pi \cdot \epsilon_\Upsilon \, p_\pi \cdot \epsilon_Z  +L^2_Z(E_\pi)\, p_\pi^2 \, \epsilon_\Upsilon  \cdot \epsilon_Z \end{array} \right)
 \nn \, ,
 \eea
where  
\bea 
L_{Z}^1(E_\pi) &=& \frac{g m_B^{3/2}}{4\sqrt{2} \pi f} \left[ -(g_\Upsilon + g_1 +3 g_2) \overline{F}(b_{BB},E_\pi + \Delta) 
+ (g_\Upsilon -2 g_2)  \overline{F}(b_{BB^*},E_\pi) \right. \\
&&\left.  +(g_2-g_1)  \overline{F}(b_{B^*B^*},E_\pi - \Delta) \right] \nn \\
L_{Z}^2(E_\pi) &=& -\frac{g m_B^{3/2}}{4\sqrt{2} \pi f} \left[ (g_\Upsilon -2 g_2) \overline{F}(b_{BB^*},E_\pi) 
+g_\Upsilon  \overline{F}(b_{B^*B^*},E_\pi-\Delta) \right ] \nn \\
L_{Z'}^1(E_\pi) &=& \frac{g m_B^{3/2}}{4\sqrt{2} \pi f} \left[ -(g_\Upsilon -2 g_2) \overline{F}(b_{BB^*},E_\pi+\Delta) 
+(g_\Upsilon +2 g_2 -2 g_1)  \overline{F}(b_{B^*B^*},E_\pi) \right ] \nn \\
L_{Z'}^2(E_\pi) &=& \frac{g m_B^{3/2}}{4\sqrt{2} \pi f} \left[ (g_\Upsilon -2 g_2) \overline{F}(b_{BB^*},E_\pi+\Delta) 
+g_\Upsilon  \overline{F}(b_{B^*B^*},E_\pi) \right ] \nn \, .
\eea
Here we have defined $\overline{F}(b,E)= F(E/b)/\sqrt{b}$.  Inserting Eq.~(\ref{loop}) into the third line of Eq.~(\ref{dressing}) 
one obtains 
\bea
i{\cal M}^{\rm loop}_Z &=& -\big( T_{ZZ}(E_B+E_{\bar B})  \, L_Z^1(E_\pi)+ T_{ZZ'}(E_B+E_{\bar B})  \,L^1_{Z'}(E_\pi) \big) p_\pi \cdot \epsilon_\Upsilon \, p_\pi \cdot \epsilon_Z 
\\
&&- \big(  T_{ZZ}(E_B+E_{\bar B})  \,L_{Z}^2(E_\pi)  + T_{ZZ'}(E_B+E_{\bar B})  \, L^2_{Z'}(E_\pi) \big)   \, p_\pi^2 \, \epsilon_\Upsilon  \cdot \epsilon_Z  \nn \\
i{\cal M}^{\rm loop}_{Z'} &=& -\big( T_{Z'Z}(E_B+E_{\bar B})  \, L_Z^1(E_\pi)  + T_{Z'Z'}(E_B+E_{\bar B})  \,L^1_{Z'}(E_\pi) \, \big) p_\pi \cdot \epsilon_\Upsilon \, p_\pi \cdot \epsilon_Z  \nn \\
&&- \big( T_{Z'Z}(E_B+E_{\bar B})  \, L_{Z}^2(E_\pi)  + T_{Z'Z'}(E_B+E_{\bar B})  \, L^2_{Z'}(E_\pi) \big)   \, p_\pi^2 \, \epsilon_\Upsilon  \cdot \epsilon_Z \nn \,. 
\eea 

For dressing the tree-level contact interactions,  we use the second line in Eq.~(\ref{dressing}). The functions $\Sigma_{B^{(*)}B^*}(E)$ have a linear 
divergence that can be removed by adding a counterterm proportional to the leading contact interaction that is being dressed. 
When this counterterm is dressed using the third line of Eq.~(\ref{dressing}), the result has the same form as the 
linear divergence in the second line in Eq.~(\ref{dressing}) and the counterterm is chosen so that the  linear divergence is removed. Alternatively, one could evaluate  $\Sigma_{B^{(*)}B^*}(E)$ in pure dimensional regularization with minimal subtraction and the linear divergence is absent.

For the amplitude for $\Upsilon(5S) \to  B^{*+}  \bar{B}^0   \pi^- $ the final result of including the loop diagrams and resumming the contact interactions 
is that $A_1^{\rm tree}$ is replaced with 
\bea
A_1(E_B,E_{\bar B},E_\pi)  &=&
A_1^{\rm tree} - \tfrac{1}{\sqrt{2}}p_\pi^2 \big( T_{ZZ}(E_B+E_{\bar B}) \, L_{Z}^2(E_\pi)  +  T_{ZZ'}(E_B+E_{\bar B}) \, L^2_{Z'}(E_\pi)  \big) \nn \\
&&-  (g_{\Upsilon \pi} + g_{\Upsilon \pi}^\prime) \frac{E_\pi}{f} \Sigma_{BB^*}(E_B+E_{\bar B}) \,T_{ZZ}(E_B+E_{\bar B}) \\
&& +   (g_{\Upsilon \pi} - g_{\Upsilon \pi}^\prime) \frac{E_\pi}{f} \Sigma_{B^*B^*}(E_B+E_{\bar B}) \,T_{ZZ'}(E_B+E_{\bar B})\nn\, , 
 \eea
 and $A_i^{\rm tree}$ is replaced with 
 \bea 
 A_i(E_B,E_{\bar B},E_\pi)  &=& A_i^{\rm tree}  -  \tfrac{1}{\sqrt{2}} \big(T_{ZZ}(E_B+E_{\bar B}) \, L^1_{Z}(E_\pi) + T_{ZZ'}(E_B+E_{\bar B}) L^1_{Z'}(E_\pi)  \big)\, ,
 \eea
for $i = 2, \ldots, 5$.  In the amplitude of the process $\Upsilon(5S) \to (B^{*+} \bar{B}^{*0})_{S=1} \pi^-$, we must make the replacements 
\bea
B_5(E_B,E_{\bar B},E_\pi) &=& B^{\rm tree}_5 -  T_{Z'Z'}(E_B+E_{\bar B})  \, L^2_{Z'}(E_\pi)  -  T_{Z'Z}(E_B+E_{\bar B})\, L_{Z}^2(E_\pi)  \\
&& + 2(g_{\Upsilon \pi} - g_{\Upsilon \pi}^\prime)\,\frac{E_\pi}{f} \Sigma_{B^*B^*}(E_B+E_{\bar B}) \,T_{Z'Z'}(E_B+E_{\bar B}) \nn \\
&&-2(g_{\Upsilon \pi} + g_{\Upsilon \pi}^\prime)\,\frac{E_\pi}{f}   \Sigma_{B B^*}(E_B+E_{\bar B}) \,T_{ZZ'}(E_B+E_{\bar B})  \nn \\
B_6(E_B,E_{\bar B},E_\pi)  &=& B_6^{\rm tree} - T_{Z'Z}(E_B+E_{\bar B}) \, L_Z^1(E_\pi)   - T_{Z'Z'}(E_B+E_{\bar B}) \, L^1_{Z'}(E_\pi)  \nn \, .
\eea
Note that ${\cal M}[\Upsilon(5S) \to (B^{*+} \bar{B}^{*0})_{S\neq1} \pi^-]$ receives no contribution from any diagram with higher order contact interactions, so is not changed upon including the loop diagrams.

The differential decay rate for $\Upsilon(5S) \to B^{*+} \bar{B}^0 \pi^-$ is given by
\bea\label{BsBpi}
\frac{d^2\Gamma[\Upsilon(5S) \to B^{*+} \bar{B}^0 \pi^-]}{dE_B dE_{\bar{B}} } &=&
\frac{m_B m_{B^*}}{192\pi^3 f^2}
\left( 3 |A_1|^2 + |A_2|^2 (p_B^2)^2 +|A_3|^2 (p_{\bar{B}}^2)^2 \right.\\
&+& (|A_4|^2 + |A_5|^2) p_B^2 \, p_{\bar{B}}^2
+2{\rm Re}[A_1^*(A_2 \,p_B^2 +A_3 \,p_{\bar{B}}^2 +(A_4+A_5) \vec{p}_B \cdot \vec{p}_{\bar B} ]   \nn \\
&+& 2{\rm Re}[A_2^* A_3+A_4^*A_5] (\vec{p}_B \cdot \vec{p}_{\bar B})^2 +2 {\rm Re}[A_2^* (A_4+A_5)] \,p_B^2 \,\vec{p}_B \cdot \vec{p}_{\bar B} \nn \\
&+&\left. 2{\rm Re}[A_3^* (A_4+A_5)] p_{\bar B}^2 \, \vec{p}_B \cdot \vec{p}_{\bar B} \right)\, .\nn
\eea
The differential decay rate for $\Upsilon(5S) \to B^{*+} \bar{B}^{*0}\pi^-$ is given by 
\bea\label{BsBspi}
\frac{d^2\Gamma[\Upsilon(5S) \to B^{*+} \bar{B}^{*0} \pi^-]}{dE_B dE_{\bar{B}} } &=&
\frac{m_{B^*}^2}{384\pi^3 f^2} \big[ 3 |B_5|^2+ 2{\rm Re}[ B_5^*B_6] p_\pi^2 + |B_6|^2 (p_\pi^2)^2 \\
&+& |B^{\rm tree}_2|^2 (3 \, p_\pi^2 (\vec{p}_B-\vec{p}_{\bar B})^2 - \, (p_B^2 - p_{\bar B}^2)^2) \nn \\
&+& |B^{\rm tree}_4|^2 \big( 28\, (p_B^2 \,p^2_{\bar B}- (\vec{p}_B \cdot \vec{p}_{\bar B})^2) + 2 \, p_\pi^2 \,(\vec{p}_B-\vec{p}_{\bar B})^2 \big) \nn \\
&+& 2  B_2^{\rm tree} B^{\rm tree}_4 \big( 4\, ( (\vec{p}_B \cdot \vec{p}_{\bar B})^2 -p_B^2 p_{\bar B}^2) + 2\, (p_B^2 - p_{\bar B}^2)^2 \big) \big] \,.\nn
\eea
Throughout Eqs.~\eqref{BsBpi} and \eqref{BsBspi} we have written $A_i$ and $B_i$ in place of $A_i(E_B,E_{\bar B},E_\pi)$ and $B_i(E_B,E_{\bar B},E_\pi)$ to make these expressions compact.

In order to apply these formulae, we need to determine the coupling constants $g_{\Upsilon \pi}$ and $g_{\Upsilon \pi}^{\prime}$ as well as the complex parameters $\gamma_+$, $\gamma_+^{\prime}$, and $\gamma_-$. Fitting the values of these parameters by fully exploring this eight (real-)dimensional space is beyond the scope of the present work.  Instead we use a hierarchical fitting procedure:  first we fit the $\gamma$ parameters using the constraints imposed by the data on $\Upsilon(5s) \to \Upsilon(nS) \pi^+ \pi^-$ and  $\Upsilon(5s) \to h_b(mP) \pi^+ \pi^-$ and then we fit $g_{\Upsilon \pi}$ and $g_{\Upsilon \pi}^{\prime}$ to reproduce the partial decay rates with the given values of the $\gamma$ parameters.

To fit the $\gamma$ parameters, we will make further simplifying assumptions. We want to fix some parameters so that the poles in the $T$ matrix agree with previous experimental and theoretical analyses and we consider three alternative schemes to do so.
\begin{itemize}
  \item Scenario (a) is to have a $T$ matrix which does not mix the $Z$ and $Z^\prime$ channels, i.e.\ taking $\gamma_- = 0$ and therefore $T_{ZZ^\prime} = 0$.  This is motivated by the empirical fact that the experimental data in Ref.~\cite{Adachi:2012cx} are fit well with only a $Z_b$ appearing in the 
$B^*\bar{B} -c.c.$ channel, and adding the $Z^{\prime}_b$ does not improve the fit. In this case  we must include HQSS violation, i.e., $\gamma_+ \neq \gamma_+^{\prime}$, to correctly produce both poles. 
Defining $\gamma_{Z^{(\prime)}} = \sqrt{M( -BE_{Z^{(\prime)}} + i \Gamma_{Z^{(\prime)}} /2)}$, where $BE_{Z^{(\prime)}}= m_{Z_b^{(\prime)}} - m_{B^{(*)}} - m_{B^*}$ and $\Gamma_{Z_b^{(\prime)}}$ is the width of the $Z_b \,(Z'_b)$, we have in this case
\bea\label{pole1}     
\gamma_+ = \gamma_Z\,, \qquad \gamma_+^{\prime} = \gamma_{Z^{\prime}} \nn \, .
\eea

  \item Scenario (b) is to take the $T$ matrix to respect HQSS and therefore to have $\gamma_+ = \gamma_+^\prime$.  Then we must have nonvanishing $\gamma_- \neq 0$ so both the $Z_b$ and $Z_b^{\prime}$ poles are correctly reproduced.
In this case, $\gamma_+$ and $\gamma_-$ are determined by the equations
\bea\label{pole2}
\gamma_+ - \gamma_Z &=& \frac{\gamma_-^2}{\gamma_+  + \sqrt{M \Delta + \gamma_Z^2}} \\
\gamma_+ - \gamma_{Z'} &=& \frac{\gamma_-^2}{\gamma_+  + \sqrt{-M \Delta +\gamma_{Z'} ^2}} \nn \, 
\eea
and $\gamma_-$ is fixed up to a sign.  We take $\Re \gamma_- > 0$.

  \item Scenario (c) is the same as Scenario (b) except we take $\Re \gamma_- < 0$.  Later we observe that this sign always gives a better fit to the data.
\end{itemize}

For each of the above three scenarios, we have to decide which data to use when we determine the location of the $Z_b$ and $Z_b^{\prime}$ poles.  In fitting to the experimental data on $\Upsilon(5S) \to B^{(*)}\bar{B}^{(*)} \pi$,  Ref.~\cite{Adachi:2012cx} determines the masses and widths of $Z_b$ and $Z_b^{\prime}$ from the 
experimental analysis of $\Upsilon(5S) \to \Upsilon(nS)\pi^+\pi^-$ and $\Upsilon(5S) \to h_b(mP)\pi^+\pi^-$, which yields $M_{Z_b} = 10607.2 \pm 1.5 \, {\rm GeV}$ and $\Gamma_{Z_b} = 11.5 \pm 2.2$ MeV, $M_{Z_b} = 10607.2 \pm 1.5 \, {\rm GeV}$ and $\Gamma_{Z_b} = 11.5 \pm 2.2$ MeV. If they try to extract these masses from the data on $\Upsilon(5S) \to B^{(*)}\bar{B}^{(*)} \pi$, they find lower masses that are 
consistent with the $Z_b$ and $Z_b^{\prime}$ being bound states.  However the errors are much larger. As emphasized in Refs.~\cite{Cleven:2013sq,Cleven:2011gp}, the location of poles 
is sensitive to the choice of line shape. Refs.~\cite{Cleven:2013sq,Cleven:2011gp} found the poles could  be below threshold if one uses their line shape, which is similar to ours. 
For our analysis, we should fit $\gamma_+$, $\gamma_+^{\prime}$, and $\gamma_-$  using data on $\Upsilon(nS)\pi^+\pi$ and $h_b(mP)\pi^+\pi^-$ since this data gives the tightest constraints 
on the parameters. Unfortunately that analysis is not available so we will try two options for fitting these parameters.
\begin{itemize}
  \item Option (1) is  demanding the poles be in the same locations as quoted in Ref.~\cite{Adachi:2012cx},
which are above threshold
  \item Option (2) is requiring the states be below threshold and have binding energies of $BE_Z = - 4.7$ MeV and $BE_{Z'} = -0.11$ MeV, as quoted in Ref.~\cite{Cleven:2011gp}.
\end{itemize}
\begin{table}[t]
{
\renewcommand\tabcolsep{6pt}
\centering\small
\begin{tabular}{lccccc}
\toprule
\textbf{Fit}\hspace*{0.15cm} & \multicolumn{5}{c}{\textbf{Parameter}} \\  \cline{2-6}
                  &  $\gamma_+$& $\gamma_+^{\prime}$& $\gamma_-$& $g_{\Upsilon \pi}  $& $g^{\prime}_{\Upsilon\pi}$ \\
\cline{1-6}
  1a &  $0.133 + 0.184\, i$ & $0.106 +0.144\, i$ & 0  & $5.8^{+2.0}_{-1.6}$  & $1.2^{+0.4}_{-0.3}$  \\
  1b & $0.110 + 0.173\, i$  & $0.110 + 0.173\, i$  & $0.100+0.005 \, i$ & $ 7.0^{+2.5}_{-3.3}$ & $ 1.72^{+0.05}_{-0.04}$  \\
  1c & $0.110 + 0.173\, i$  & $0.110 + 0.173\, i$  & $ -0.100 - 0.005 \, i$ & $4.6^{+1.4 }_{-1.9}$ & $ 1.0^{+0.5}_{-0.4 }$  \\
  2a & $0.200+0.122\, i$  & $0.125 +0.122\, i$ & 0 & $5.5^{+1.9}_{-2.6}$ &  $0.8^{+0.6}_{-0.4}$ \\
  2b &  $0.162 + 0.142\, i$ & $0.162 + 0.142\, i$ & $ 0.118 - 0.045 \, i$ &  $6.4^{+2.3}_{-3.2}$ &  $1.7^{+0.06}_{-0.1}$ \\
  2c &  $0.162 + 0.142\, i$ & $0.162 + 0.142\, i$ & $ -0.118 + 0.045 \, i$ &  $4.0^{+1.1}_{-1.5}$ &  $0.1^{+0.8}_{-0.6}$ \\
\botrule
\end{tabular}}
\caption{Parameters for six fits discussed in the text. $\gamma_+$, $\gamma_+^{\prime}$ and $\gamma_-$ are in units of GeV,
$g_{\Upsilon \pi}$ and $g_{\Upsilon \pi}'$ are in unites of GeV${}^{-5/2}$. }
\label{table}
\vspace{0.25 in}
\end{table}

Once the $\gamma$ parameters are fit, the only remaining undetermined parameters are $g_{\Upsilon\pi}$ and $g'_{\Upsilon\pi}$. These always appear in the linear combinations
$g_{\Upsilon\pi} \pm g'_{\Upsilon\pi}$. We determine these couplings by requiring that we reproduce the correct rates 
for $\Upsilon(5S) \to B \bar{B}^* \pi, B^* \bar{B} \pi$  and $\Upsilon(5S) \to B^* \bar{B}^* \pi$.
Combining the total width from the PDG and the branching fractions recently measured in Ref.~\cite{Adachi:2012cx}, we obtain 
\bea\label{totalrates}
\Gamma[\Upsilon(5S) \to B \bar{B}^* \pi] + \Gamma[\Upsilon(5S) \to  B^* \bar{B} \pi] &=& 2.3 \pm 1.2 \, {\rm MeV} \\
\Gamma[\Upsilon(5S) \to B^* \bar{B}^* \pi] &=& 1.2 \pm 0.6 \,{\rm MeV} \, .\nn
\eea
Here we have combined all quoted errors in quadrature. We compute these rates by summing over all channels using Eqs.~(\ref{BsBpi},\ref{BsBspi}) with neutral channels multiplied by a factor of $1/2$ and a common isospin averaged pion mass of $138$ MeV. The results for all combinations of the three scenarios and two options for the $\gamma$ parameters are shown in Table~\ref{table}.  The errors shown in the are estimated by varying the rates in Eq.~(\ref{totalrates}) between their high and low values. Note that the dominant uncertainty in Eq.~(\ref{totalrates}) is due to the uncertainty in the total width of the $\Upsilon(5S)$ quoted in the PDG, not the branching ratios, so the errors in Eq.~(\ref{totalrates}) are highly correlated. Note that in all of our fits $g_{\Upsilon \pi}^\prime  \ll g_{\Upsilon \pi}$ which is consistent with  HQSS.

\begin{figure}[t]
\begin{center}
\includegraphics[width=6.6in]{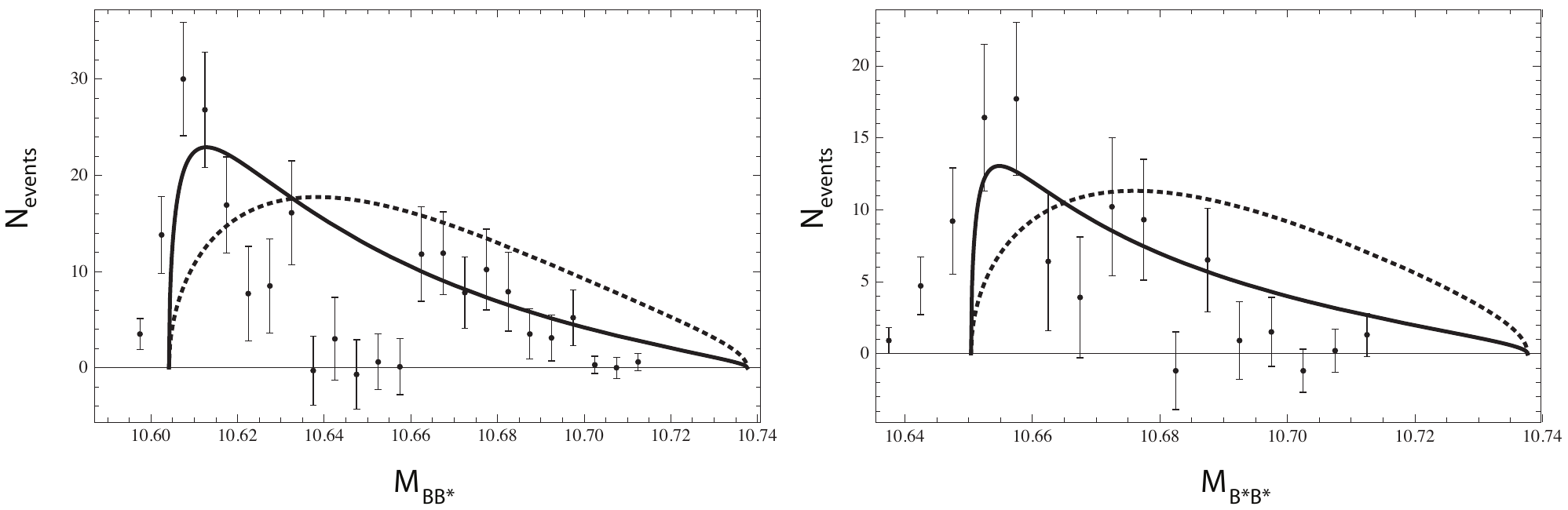}
\caption{Number of events as a function of the the invariant mass of the final state $B$ mesons in $\Upsilon(5S) \to B\bar{B}^* - c.c.$ (left) and 
$\Upsilon(5S) \to B^*\bar{B}^*$ (right). The data is from Ref.~\cite{Adachi:2012cx} and have had background subtracted. The solid (dashed)  line is the full (tree-level) 
calculation of the invariant mass distribution multiplied by an arbitrary normalization. The parameters used are from Fit 1a.}
\label{Fit1a}
\end{center}
\end{figure}

\begin{figure}[t]
\begin{center}
\includegraphics[width=6.6in]{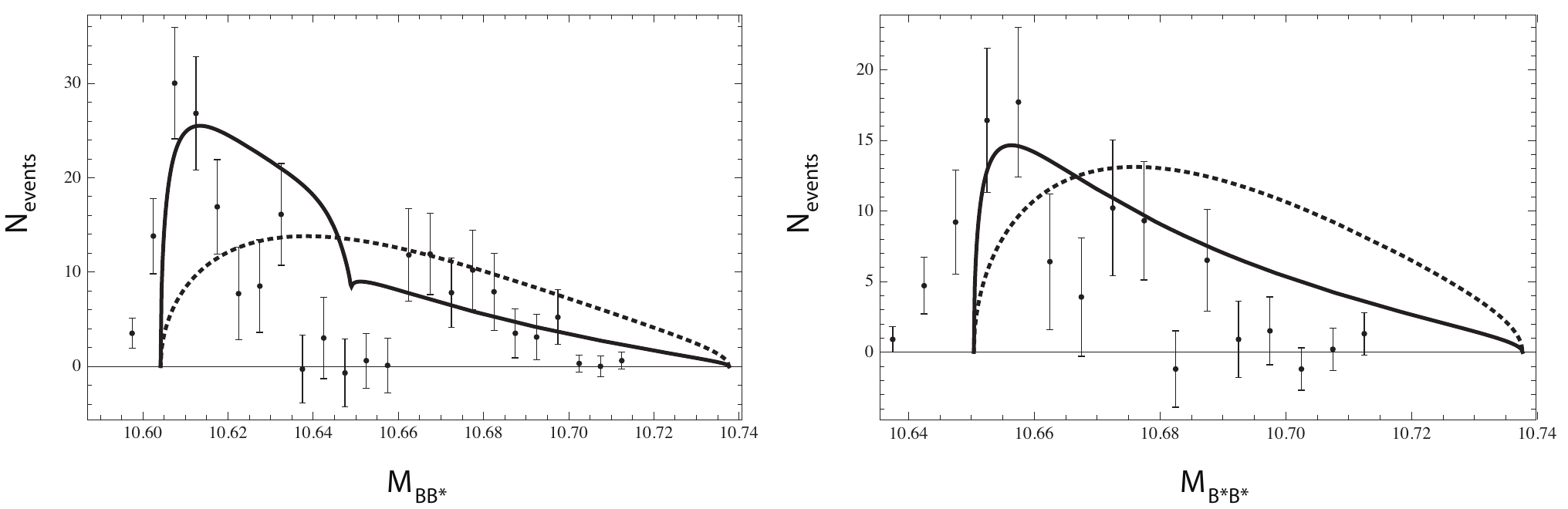}
\caption{Number of events as a function of the the invariant mass of the final state $B$ mesons in $\Upsilon(5S) \to B\bar{B}^* - c.c.$ (left) and 
$\Upsilon(5S) \to B^*\bar{B}^*$ (right). The data is from Ref.~\cite{Adachi:2012cx} and have had background subtracted. The solid (dashed)  line is the full (tree-level) 
calculation of the invariant mass distribution multiplied by an arbitrary normalization. The parameters used are from Fit 1c.}
\label{Fit1c}
\end{center}
\end{figure}

The resulting distributions  as a function of $m_{BB^*}$ or $m_{B^*B^*}$  for the cases 1a and 1c  are shown in Fig.~\ref{Fit1a} (1a) and 
Fig.~\ref{Fit1c} (1c). The solid line is the full calculation, the dotted line is the result if only tree-level diagrams are kept. 
The data are number of events so we have multiplied both differential distributions by an arbitrary normalization chosen to agree with data.  The first thing to point out is that the theoretical curves vanish at the correct thresholds $m_{BB^*} = m_B + m_{B^*} = 10.604$ GeV
and $m_{BB^*} = 2 m_{B^*} = 10.650$ GeV. The data is nonvanishing below these thresholds. This is probably related to experimental resolution 
 and our calculation  needs to be convolved with a smearing function to make a sensible comparison with data.\footnote{We thank R. Mizuk for a discussion on this point.} We also should convolve the differential rate with a Breit-Wigner reflecting the fact that the $\Upsilon(5S)$ has a finite width. Because of these 
issues we choose not to fit our parameters to the experimental data in these plots.  

The predicted distributions are nearly identical for Fits 1a, 1b, 1c and 2a, 2b, 2c, respectively. That is, the distributions have very similar shapes for 
the two choices of the location of the $Z_b$ and $Z_b^\prime$ poles. The fits 1b and 2b yield a curve which shows a peak due to the $Z_b^{\prime}$ in the $B\bar{B}^* - c.c$ channel in the mass range $10. 64 \, {\rm GeV}< M_{BB*} < 10.66 \, {\rm GeV}$ where the number of events vanishes. These distributions  are in qualitative disagreement with the data so we do not show plots of the distributions for these choices of parameters. The fits 1a and 2a yields curves which do not reproduce this dip but are in qualitative agreement on either side of the dip. In the  fits 1c and 2c the effect of $Z_b^{\prime}$ is to suppress the $B^* \bar{B} - c.c.$ channel cross section  in the region where there are no events. 
The plots in Figs.~\ref{Fit1a} and \ref{Fit1c} clearly show that resumming the final-state interactions improves the agreement  with data relative to the tree-level calculation. In particular, the peaks 
in our distributions are in the correct locations. When more precise data on these distributions becomes available, it would be interesting to 
fit the parameters of our theory directly to the line shapes to see if we can reproduce some of the finer structure. This would require taking into 
account effects due to the width of the $\Upsilon(5S)$ as well as experimental resolution.

\section{Angular Distributions in  $\Upsilon(5S) \to Z_b^{(\prime)}\pi^{\pm}$}

In this section, we will focus on  the $\Upsilon(5S) \rightarrow Z_b^{(\prime)+} \pi^-$ transition at the $m_{B^*} + m_{B^{(*)}}$ threshold.
At this kinematic point we have $\vec{p}_B = \vec{p}_{\bar B} = -\vec{p}_\pi/2$. After summing over the polarization of the $Z_b^{(\prime)}$, the matrix elements squared can   be written as
\bea\label{msqr}
|\calM[\Upsilon(5S) \rightarrow Z_b^{(\prime)+}\pi^-] |^2 = P_{Z^{(\prime)}} |\epsilon_\Upsilon \cdot \hat{p}_\pi|^2  + T_{Z^{(\prime)}}
 |\epsilon_\Upsilon \times\hat{p}_\pi|^2 \, ,
\eea
where the coefficients $P_{Z^{(\prime)}}$ and  $T_{Z^{(\prime)}}$ are given by
\bea
P_Z &=& |A_1 +(A_2+A_3+A_4+A_5)p_\pi^2/4|^2 \\ 
T_Z &=& |A_1|^2\nn \\
P_{Z'} &=& |B_5 + B_6 p_\pi^2|^2/2\nn \\
T_{Z'} &=& |B_5|^2/2 \, , \nn 
\eea
and we have again dropped the arguments in $A_i(E_B,E_{\bar B},E_\pi)$ and  $B_i(E_B,E_{\bar B},E_\pi)$ to make these expressions compact. Since we require
the $B$ and ${\bar B}$ mesons to be at threshold we must evaluate these expressions at $E_B = E_{\bar B}$ and $E_\pi = E_{\pi,{\rm max}}$.    Since the $\Upsilon(5S)$ is produced in $e^+e^-$ collisions with polarization transverse to the beam, the angular distribution of the pion relative to the beam axis
can be nontrivial. Defining the angle the pion makes with the beam to be $\cos \theta$, the angular distribution is
\bea
\frac{d\sigma}{d\Omega} \propto 1 + \rho_{Z^{(\prime)}} \cos^2\theta \, 
\eea
where 
\bea
\rho_{Z^{(\prime)}} = \frac{T_{Z^{(\prime)}} - P_{Z^{(\prime)}} }{   T_{Z^{(\prime)}}+P_{Z^{(\prime)}}   }\, .
\eea 
Similar angular distributions were studied in $X(3872)$ production and decay in Refs.~\cite{Mehen:2011ds,Margaryan:2013tta}.  If $P_{Z^{(\prime)}} = T_{Z^{(\prime)}}$ the angular distribution becomes uniform. One can see from the amplitudes that this is case for the diagrams in which the pion is produced from one of the contact interactions. Values of $ \rho_{Z^{(\prime)}}$ different from zero come from the diagrams in which the pion couples directly to the 
$B$ mesons. This would be all the diagrams in Fig.~\ref{oneloop} or all diagrams but the ones on the left in Fig.~\ref{fig:UpsilonZb} and Fig.~\ref{eqn:UpsilonZbpPi}.
Thus the variables $\rho_{Z^{(\prime)}}$  provide a means of distinguishing between the production mechanisms for the $Z_b$ and $Z_b^\prime$.
From inspecting amplitudes one can also verify that the $\rho_{Z^{(\prime)}}$ parameters vanish in the heavy quark limit, so they are expected to be small.
In fact, in order to produce the observed total rates, we find that our extracted values for the couplings   $g_{\Upsilon \pi}$ and $g_{\Upsilon \pi}^\prime$ are numerically large 
relative to the couplings of the $\Upsilon(5S)$ to $B^{(*)}\bar{B}^{(*)}$. So the contact interactions dominate the decay rate and 
the parameters $\rho_Z$ and $\rho_{Z'}$ are further suppressed. The value of $\rho_Z$ we find depends on the fit: $\rho_Z = 0.016, 0.026, 0.008, 0.013, 0.031, 0.001$ in Fits  1a, 1b, 1c, 2a, 2b,and 2c, respectively.    Curiously, $\rho_{Z'} = -0.021$ or $-0.022$ in all six fits. In all cases the magnitude of $\rho_Z$ and  $\rho_{Z'}$ is order a few percent or smaller, and therefore will be difficult to  distinguish from $\rho_{Z^{(\prime)}} =0$. It would be interesting to explore how the parameters 
$\rho_{Z^{(\prime)}}$ depend on the energy of the pion  but we expect them to continue to be at the few percent level throughout phase space and so we will not study this further in this paper.

\section{Conclusions}

In this paper we have  computed the distributions in $\Upsilon(5S) \to B^{(*)} \bar{B}^{(*)} \pi$ using an effective field theory for 
strongly interacting $B$ mesons near threshold. We first fixed some couplings of $\Upsilon(5S) \to B^{(*)}\bar{B}^{(*)}$ using available data on these decays and found HQSS violating operators are needed for consistency with available data. We then analyzed $\Upsilon(5S) \to B^{(*)} \bar{B}^{(*)} \pi$ and find that the decay rate is dominated by contact interactions that couple the $\Upsilon(5S)$, $B^{(*)}$ and $\bar{B}^{(*)}$ mesons, and the pion. The relative size of the extracted contact interactions are consistent with HQSS.  Resumming final state interactions of the strongly interacting $B$ mesons after the pion is emitted 
produces line shapes that are in qualitative agreement with data. There are several directions one could pursue following this analysis.  It would be interesting to repeat the analysis of $\Upsilon(5S) \to \Upsilon(nS) \pi^+ \pi^-$
and $\Upsilon(5S) \to h_b(mP) \pi^+ \pi^-$ using the line shapes in this paper and compare with the results of Refs.~\cite{Cleven:2011gp,Cleven:2013sq}.
 It would also be interesting to incorporate range corrections into the $T$-matrices in Eq.~(\ref{Tmatrix}). This would introduce terms linear in the energy in the denominators of the $T$-matrices, yielding line shapes that are more similar to the one used in  Refs.~\cite{Cleven:2011gp,Cleven:2013sq}. Finally, it would be useful to fit data simultaneously on  $\Upsilon(5S) \to \Upsilon(nS) \pi^+ \pi^-$, $\Upsilon(5S) \to h_b(mP) \pi^+ \pi^-$, and $\Upsilon(5S) \to B^{(*)} \bar{B}^{(*)} \pi$, all computed within the same theoretical framework,  to constrain the parameters in the $T$-matrices. Such an analysis could help determine the location of  the  $Z_b$ and $Z_b^\prime$ poles and aid in the interpretation of the $Z_b$ and $Z_b^\prime$ states.

\acknowledgments 

We thank R. Mizuk for correspondence related to this work. This work was supported in part by the  Director, Office of Science, Office of High Energy
Physics, of the U.S. Department of Energy under Contract No.   
DE-FG02-05ER41368.



\end{document}